# Quantum Model for Electro-Optical Phase Modulation

**José Capmany[1,*] and Carlos R. Fernández-Pousa[2]**

[1] ITEAM Research Institute, Universidad Politécnica de Valencia, 46022 Valencia, Spain

[2] Signal Theory and Communications, Dep. of Physics and Computer Science, Univ. Miguel Hernández, 03202 Elche, Spain

[*]Corresponding author: jcapmany@iteam.upv.es

We present a detailed analysis of the quantum description of electro-optical phase modulation. The results define a black-box type model for this device which may be especially useful in the engineering steps leading to the design of complex quantum information systems incorporating one or more of these devices. By constructing an explicit representation of the phase modulation scattering operator, it is shown that an approach based entirely on its classical description leads to unphysical modes associated to non-positive frequencies. After modifying this operator, phase modulation is described, for the first time to the best of our knowledge, in terms of a unitary scattering operator $\hat{S}$ defined over positive-frequency modes. The modifications introduced by $\hat{S}$ in the process of sideband generation by phase modulation are shown to be not significant when the radiation belongs to optical bands, thus being consistent with the classical description. Finally, the model is employed to characterize the important case of multitone radiofrequency modulation of an optical signal.

# 1. Introduction

Phase modulation constitutes a basic technology for manipulating optical waves at both classical and quantum levels. At the quantum side, phase modulators are now being used in a variety of quantum applications, including sources of approximated single-photon states in quantum key distribution (QKD) systems [1], frequency-coded [2, 3] and subcarrier multiplexed [4] QKD systems or, more recently, for tailoring the wave function of heralded photons [5]. Furthermore, due to their inherent capabilities together with its ease of integration, it is envisaged that its use will expand both to more complex QKD configurations as well as in other emerging applications in the field of quantum information systems.

The development and practical implementation of the later requires, at a certain step, not only a physical knowledge of the underlying physical principles but also a considerable engineering design step for which blackbox-type models of the essential photonic components are required. These models are available at the quantum level for several classical passive optical devices [6], but it has only been recently that the theory of quantum electro-optic modulation has gained some attention [7, 8].

Modulators are nonlinear devices producing a multimode output, observed as the multiple sidebands that a single optical carrier develops after tone modulation. From the quantum point of view, the sideband generation can be interpreted as the generation of a multimode output state from a singlemode input by means of a scattering device. A quantum theory should describe such a modulation regardless the values of the photon frequencies involved, even if they are away from the typical optical frequencies where a well-established classical theory is at hand.

In this regard, several effects may show up in this process of sideband or multimode state generation, particularly when the sidebands depart substantially from the optical carrier. As more distant sidebands are generated, dispersion tends to mismatch the phase of the waves inside the modulator and decrease the modulator's bandwidth [9]. In the classical description, this effect is taken into account by considering that the modulation index $m$ attained by the modulator at a certain voltage level depends on the modulating radio-frequency tone $\Omega$. The problem is then reduced to the description of the modulator for arbitrary values of $\Omega$ and $m(\Omega)$: multitone modulation is then associated to a combined modulation of several tones, each with its modulation index. This is precisely the route followed in this paper.

A second effect is related to the observation that, as frequencies becomes larger, any guided optical medium ceases to be a single-mode guide and, as frequencies become lower, asymmetric guides cease to be guiding. The guided or radiated character of the modes associated to the output frequencies depends, of course, on the characteristics of the modulator. But leaving aside the output spatial distribution, the problem can still be posed as the determination of input-output relations between states with definite frequency mediated by a scatterer which tends to increase or decrease the input mode frequency by multiples of an amount $\Omega$.

There is, however, a more fundamental issue that cannot be circumvented by the assumption of an ideal device where the aforementioned effects are absent. In the classical description of phase modulators (PM) the number of output modes is always infinity leading to an unavoidable difficulty at the quantum level, since the strict phase modulation of a carrier will always develop unphysical negative-frequency modes, even classically. As will be shown in this paper, the presence of these non-physical modes is necessary for the unitarity of the scattering operator, and therefore its exclusion renders the theory non-unitary. The objective of this paper is precisely to overcome this fundamental problem by constructing a phase-modulation multimode scattering operator which is unitary in the Hilbert space of positive-frequency modes.

The paper is organized as follows. In section 2 we briefly review the equations of the classical phase modulator. The purpose of this section is twofold. Firstly, any quantum description of the PM operation must be consistent with the classical model, so the equations are written for later reference. Secondly, the classical model serves as a starting point when trying a quantum description of PM using the base of coherent states. In section 3 we develop such a classical-like PM scattering theory. After a brief qualitative description of the scattering process in section 3.1, we provide in section 3.2 a first (tentative) description of a PM operator $\hat{S}$ using the overcomplete base of coherent states and a proper analogy with the classical PM operation, observing that, since within this approach there is no restriction on the sign (positive or negative) of the frequencies of the involved modes, the quantum theory is unitary only in a Hilbert space containing unphysical modes. From that we obtain the required transformation equations for the mode creation and annihilation operators under $\hat{S}$, which are expressed as a linear canonical transformation. In section 3.3 we construct a representation of this (tentative) unitary PM scattering operator $\hat{S}$ and in section 3.4 we introduce a diagrammatic procedure for computing its action.

In order to avoid the generation of unphysical modes with non-positive frequencies, the representation of the scattering operator is modified in section 4.1, yielding a new operator $\hat{S}$ which is unitary in the space of positive-frequency modes. The explicit form of the transformation equations for the mode creation operators is then derived in section 4.2 using the diagrammatic method. Finally, in section 5 we consider the case of multitone phase modulation which is becoming increasingly important in practice, especially in the context of subcarrier multiplexed QKD systems [10]. To our knowledge and, as opposed to other prior works [7, 8], this is the first time that such an explicit form of the PM scattering operator is reported.

## 2. Classical Operation of the Phase modulator

We briefly recall in this section the operational principles of the electro-optic phase modulator under classical regime [11], since they will be useful when comparing with the equations describing the operation of the device under quantum regime. The typical configuration of a waveguide electro-optic modulator is depicted in the upper part of figure 1. It consists of a dielectric waveguide and two electrodes placed at both sides. The dielectric material is subject to the electro-optic effect and the voltage applied to the electrodes changes linearly the refractive index undergone by one of the two possible input linear polarizations (we take for simplicity the $\hat{\mathbf{x}}$ polarization) while not affecting the refractive index in the other linear polarization (the $\hat{\mathbf{y}}$). If $\eta_0$ represents the refractive index experienced by the input $\hat{\mathbf{x}}$ polarization when no voltage is applied and $l$ represents the device length, then the extra phase change undergone by an input $\hat{\mathbf{x}}$ polarization when we apply an input voltage signal $V(t) = V_{DC} + \Delta V(t)$ is given by:

$$\Delta\phi(t) = \frac{2\pi\Delta\eta}{\lambda_0}l = \frac{\pi\eta_0^3 r}{\lambda_0}\frac{l}{d}(V_{DC} + \Delta V(t)) = \pi\frac{V_{DC} + \Delta V(t)}{V_\pi}. \tag{1}$$

In the above equation, $r$ represents the relevant electro-optic coefficient for the input $\hat{\mathbf{x}}$ polarization, $\lambda_0 = 2\pi c/\omega_0$ is the wavelength in vacuum of the input signal to the modulator, $V_{DC}$ and $\Delta V(t)$ represents the DC bias voltage and the time varying modulation signal applied to the modulator electrodes, respectively, $d$ is the distance between electrodes, and $V_\pi = \lambda_0 d/\eta_0^3 rl$.

The operation of the modulator can be described as follows. Input light is assumed monochromatic at frequency $\upsilon_0$ and propagates in the positive $z$ direction with propagation constant $k(\omega_0) = 2\pi\eta_0/\lambda_0 = \eta_0\omega_0/c \equiv \omega_0/v > 0$, where $v$ is the speed of light in the medium, which will be assumed dispersion-free for simplicity. We also assume that input light is polarized in the $\hat{\mathbf{x}}$ direction with amplitude given by $E_0$:

$$\mathbf{E}_{in}(z,t) = E_0 \hat{\mathbf{x}} e^{j(\omega_o t - k(\omega_o)z)} = E_0 \hat{\mathbf{x}} e^{j\omega_o(t-z/v)}. \tag{2}$$

The modulator is located between $z = -l$ and $z = 0$. At the output ($z = 0$) of the PM the wave acquires an additional phase:

$$\mathbf{E}_{out}(0,t) = \mathbf{E}_{in}(-l,t)e^{j\Delta\phi(t)} = E_0 \hat{\mathbf{x}} e^{j\omega_o t} e^{j\varphi_b} \exp(j\pi \Delta V(t)/V_\pi), \tag{3}$$

where $\varphi_b = k(\omega_0)l + \pi V_{DC}/V_\pi$. We will be specifically interested in the case of a sinusoidal modulation of frequency $\Omega$ given by:

$$\Delta V(t) = V_m \cos(\Omega t + \theta), \tag{4}$$

in which case, and defining the modulation index as $m = \pi V_m/V_\pi$, Eq. (3) transforms to:

$$\mathbf{E}_{out}(0,t) = \mathbf{E}_{in}(-l,t)e^{j\Delta\phi(t)} = E_0 \hat{\mathbf{x}} e^{j\omega_o t} e^{j\varphi_b} e^{jm\cos(\Omega t + \theta)}. \tag{5}$$

We can further develop equation (5) to show that the modulation process not only produces the desired fundamental tone appearing as two sidebands separated from the optical carrier by $\pm\Omega$ but also the generation of harmonics, by using the Jacobi-Anger expansion in terms of the Bessel functions of first kind:

$$e^{jz\cos\vartheta} = \sum_{n=-\infty}^{\infty} j^n J_n(z) e^{jn\vartheta}, \tag{6}$$

so that after the modulator ($z > 0$) the wave is composed of a number of travelling waves with different frequencies,

$$\mathbf{E}_{out}(z,t) = E_0 \hat{x} e^{j\varphi_b} \sum_{n=-\infty}^{\infty} (je^{j\theta})^n J_n(m) e^{j(\omega_o + n\Omega)(t-z/v)} = E_0 \hat{x} \sum_{n=-\infty}^{\infty} C_n e^{j(\omega_o + n\Omega)(t-z/v)}, \tag{7}$$

where we have defined the coefficients:

$$C_n = e^{j\varphi_b} (je^{j\theta})^n J_n(m). \tag{8}$$

Note that the inverse operation to that described by (5) is given by:

$$\begin{aligned}\mathbf{E}_{in}(-l,t) &= \mathbf{E}_{out}(0,t) e^{-j\varphi_b} e^{-jm\cos(\Omega t + \theta)} = \mathbf{E}_{out}(0,t) \sum_{n=-\infty}^{\infty} C_n^* e^{-jn\Omega t} \\ &= \mathbf{E}_{out}(0,t) \sum_{n=-\infty}^{\infty} C_{-n}^* e^{jn\Omega t} = \mathbf{E}_{out}(0,t) \sum_{n=-\infty}^{\infty} \tilde{C}_n e^{jn\Omega t},\end{aligned} \tag{9}$$

where the coefficients for the inverse transformation are $\tilde{C}_n = C_{-n}^*$. Note finally that, if the input signal is multimode and comprises frequencies given by $\omega_0 + n\Omega$ then:

$$\mathbf{E}_{in}(-l,t) = E_0 \hat{x} e^{j\omega_o t} \sum_{n=-\infty}^{\infty} \alpha_n e^{jn\Omega t} \tag{10}$$

and we can write the output travelling field exiting the modulator as:

$$\mathbf{E}_{out}(z,t) = E_0 \hat{x} \sum_{q,n=-\infty}^{\infty} \alpha_n C_{q-n} e^{j(\omega_o + q\Omega)(t-z/v)}, \tag{11}$$

which means that the multimode field is reordered by the action of the PM.

Finally, we point out the inconsistency in this classical description of phase modulation since, as it is apparent from (7), for sufficiently negative values of the index $n$, frequencies become zero or negative. However, under practical operation conditions, $\omega_0/\Omega \approx 10^4$ for optical PM and, in addition, and according to Carson's rule [12], the amplitude of the generated sidebands $C_q \sim J_q(m)$ becomes progressively negligible for $|q| > m+1$, which implies that the possibility of generating negative or zero frequencies is of little concern. In a quantum theory, however, the interpretation of these non-positive frequencies is to be analyzed in detail because all modes, even if their probability of practical generation is negligible, contribute to the unitarity of the theory at the same footing.

When dealing with a time-dependent, phase modulated real signal the existence of negative frequencies is not a severe problem because the symmetry $\cos(\omega t) = \cos(-\omega t)$ always renders negative frequencies positive [12]. Such a procedure can in principle be applied to (7), since the electric field is a real quantity. However, if we reverse the argument of a travelling-wave field $\cos(\omega t - kz)$ when the frequency becomes negative, the sign of the wave vector changes accordingly, so that it would represent a travelling wave *entering* the modulator from the right. A different possibility could be reversing the sign of both frequency and wave vector, as is suggested by our formula (7). This is again unsatisfactory, since the physical process behind is simply that incoming radiation at a sufficiently low frequency, say $\omega$, cannot decrease its value by $\Omega$, only increase it by this amount: there is no new wave at the 'folded' frequency $\Omega - \omega > 0$. This observation thus represents an inconsistency when trying to translate the classical picture to a quantum theory. Before facing this problem we will formalize in the following section the quantum theory with these non-positive frequencies, as a preliminary step towards our solution.

## 3. Quantum Phase Modulation with Non-Positive Frequencies

### 3.1 Quantum scattering through a phase modulator

As suggested by equation (7) a quantum model for electro-optic phase modulation can be formulated as a one-dimensional problem where the modulator acts a scattering region. Incoming continuous wave (CW) radiation undergoes no reflection, and increases or decreases its frequency by exchanging energy with the external radio-frequency field through their interaction with the electro-optic dielectric, thus describing an inelastic and reflectionless scattering process. The scattering is fully determined by considering radiation at arbitrary frequency, incoming from both sides of the scatterer. Equation (7) is simply the classical expression of the scattering where the CW field impinges the modulator from the left, and a similar expression would describe the action of the modulator when radiation enters the modulator from the right. In this case, nevertheless, the coefficients (8) may change depending on the internal architecture of the device. For instance, the electrodes in travelling-wave modulators are transmission lines where the radio-frequency is phase-matched to the incoming optical radiation in order to increase bandwidth [9]. Such a phase matching can be realized only for one direction of propagation, so that the parameters characterizing the same modulator when operated in reversed form are generally different. In quantum-mechanical terms this simply reflects the lack of parity invariance of the scatterer.

In order to formulate these observations we assume that the scattering problem is defined in a one-dimensional geometry with quantization length $L$. Then, the travelling modes $\exp[j(\omega t - kz)]$ allowed by periodic boundary conditions are those given by $kL = 2\pi n$, with $n$ a non-zero integer, $n = \pm 1, \pm 2, \ldots$ The corresponding frequency is $\omega = c|k| = 2\pi |n| c/L$. Modes with $n > 0$ (resp. $n < 0$) represent waves travelling to the right (resp. to the left). When the phase modulator is inserted in this quantization length, the absence of reflection implies that incoming right-moving modes transform into outgoing right-moving modes, as described classically by eq. (7). The scattering is thus fully described by an operator $\hat{S}$ transforming right-moving (with $n>0$) modes into right-moving modes, together with an operator $\hat{S}'$ acting only on left-moving modes. Although these two operators possibly depend on different parameters, they should be of the same form, and so we will restrict our presentation to the first of them.

The restriction to right-moving modes with $n>0$ permits that radiation at a given wave vector $k = 2\pi n_0/L > 0$ can be unambiguously labeled by its frequency $\omega_0 = 2\pi n_0 c/L$. We may label modes by its frequency, so that the state $|1\rangle_{\omega_0}$ stands for a one-photon state at frequency $\omega_0$. Alternatively, we use the equivalent notation $|1\rangle_{n_0}$, where the frequency is replaced by its integer index. Correspondingly, the modulation frequency $\Omega = 2\pi Nc/L$ will be labeled by an integer $N>0$. With this notation, the non-positive frequency problem is associated to states $|\Omega\rangle_n$ whose frequency index $n$ is zero or negative.

## *3.2 Description of PM via coherent states*

We use the basis of coherent states to describe the action of the electro-optic phase modulator. Referring to Figure 1 and by analogy with (7) we define the action of the scattering operator $\hat{S}$ describing the PM for a singlemode input:

$$|\Psi_{in}\rangle = |\alpha\rangle_{n_0} \xrightarrow{\hat{S}} |\Psi_{out}\rangle = \hat{S}|\alpha\rangle_{n_0} = \bigotimes_{q=-\infty}^{\infty} |C_q \alpha\rangle_{n_0+qN}. \tag{12}$$

Leaving temporarily aside the problem of negative-frequency modes, which will be tackled in section 4, we proceed by showing the unitarity of the resulting theory. Indeed, we can extend (12) to the case of an arbitrary multimode coherent input $|\{\alpha\}\rangle$, by analogy with Eq. (11):

$$|\{\alpha\}\rangle \equiv \bigotimes_{q=-\infty}^{\infty} |\alpha_q\rangle_{n_0+qN} \xrightarrow{\hat{S}} \bigotimes_{q=-\infty}^{\infty} \left| \sum_{k=-\infty}^{\infty} C_{q-k}\alpha_k \right\rangle_{n_0+qN}. \tag{13}$$

To describe the operation of the inverse scattering operator $\hat{S}^{+1}$ we have, by analogy with (7):

$$|\{\alpha\}\rangle = \bigotimes_{q=-\infty}^{\infty} |\alpha_q\rangle_{n_0+qN} \xrightarrow{\hat{S}^{-1}} \bigotimes_{q=-\infty}^{\infty} \left|\sum_{k=-\infty}^{\infty} \tilde{C}_{q-k}\alpha_k\right\rangle_{n_0+qN} = \bigotimes_{q=-\infty}^{\infty} \left|\sum_{k=-\infty}^{\infty} C^*_{k-q}\alpha_k\right\rangle_{n_0+qN}. \quad (14)$$

To prove the unitarity of $\hat{S}$, we first show that it holds for coherent states,

$$\langle\{\beta\}|\hat{S}|\{\alpha\}\rangle^* = \prod_q {}_{n_0+qN}\left\langle \beta_q \left| \sum_k C_{q-k}\alpha_k \right.\right\rangle^*_{n_0+qN} =$$

$$= \exp\left[-\frac{1}{2}\sum_q \left|\sum_s C_{q-s}\alpha_s\right|^2 - \frac{1}{2}\sum_q |\beta_q|^2 + \sum_q \sum_s \alpha^*_s C^*_{q-s}\beta_q \right] =$$

$$= \prod_q \exp\left[-\frac{1}{2}\left|\sum_n \tilde{C}_{q-n}\beta_n\right|^2 - \frac{1}{2}\sum_q |\alpha_q|^2 + \sum_s \alpha^*_q \tilde{C}_{q-s}\beta_s \right] =$$

$$= \prod_q {}_{n_0+qN}\left\langle \alpha_q \left| \sum_s \tilde{C}_{q-s}\beta_s \right.\right\rangle_{n_0+qN} = \langle\{\alpha\}|\hat{S}^{-1}|\{\beta\}\rangle, \quad (15)$$

where implicit sums and products from $-\infty$ to $\infty$ have been assumed and we have used that $|\sum_n C_{q-n}\alpha_n|^2 = |\alpha_q|^2$ and $|\sum_n \tilde{C}_{q-n}\beta_n|^2 = |\beta_q|^2$. Now, the generalization to any arbitrary state is straightforward after the use the over-complete basis of coherent states. Furthermore, using the closure relation for the basis of coherent states we obtain the following transformations of the mode creation operators:

$$\hat{S}\hat{a}^\dagger_n\hat{S}^\dagger = \sum_{q=-\infty}^{+\infty} C_q \hat{a}^\dagger_{n+qN} \qquad \hat{S}^\dagger\hat{a}^\dagger_n\hat{S} = \sum_{q=-\infty}^{+\infty} C^*_{-q} \hat{a}^\dagger_{n+qN}, \quad (16)$$

which mean that both the phase modulation operator $\hat{S}$ and its inverse $\hat{S}^\dagger$ can be interpreted as performing a linear canonical transformation. From (16) it is straightforward to compute the modulator's response to a single photon input: bearing in mind that the vacuum state $|vac\rangle$ can be identified with the multimode $|\{\alpha=0\}\rangle$ coherent state, then, according to (14) we have $\hat{S}^\dagger|vac\rangle = |vac\rangle$. Moreover, since $\hat{a}_n^\dagger|vac\rangle = |1\rangle_n$ we get [7]:

$$\hat{S}\hat{a}_n^\dagger\hat{S}^\dagger|vac\rangle = \hat{S}\hat{a}_n^\dagger|vac\rangle = \hat{S}|1\rangle_n = \sum_{q=-\infty}^{\infty} C_q \hat{a}_{n+qN}^\dagger|vac\rangle = \sum_{q=-\infty}^{\infty} C_q |1\rangle_{n+qN}. \qquad (17)$$

The coefficient $C_q$ thus has the interpretation of a probability amplitude for a transition from a one-photon state at mode $n$ to its $q$-th sideband, ie., to mode $n+qN$. Furthermore, it can be easily shown that the number of photons $\hat{N}_{ph}$ is conserved $[\hat{N}_{ph}, \hat{S}] = 0$.

## 3.3 Scattering operator

The description developed in the prior section is useful in determining the defining properties of the scattering operator associated to the quantum operation of the electro-optic phase modulation but it does not render an explicit form of such operator. In this section we provide such a form, for the first time to the best of our knowledge. After its definition we demonstrate that it provides equivalent expressions to those derived in the previous section. This form of the scattering operator will be modified in what follows to tackle with the negative frequencies. Explicitly, the scattering operator $\hat{S}$ is given by the following expression:

$$\hat{S} \equiv \hat{S}_N(\chi, \varphi_b) = \exp\left[j\hat{G}_N(\chi, \varphi_b)\right] = \exp\left[j(\chi\hat{T}_N + \chi^*\hat{T}_N^\dagger + \varphi_b\hat{N}_{ph})\right] \qquad (18)$$

where $\hat{T}_N = \sum_{n=-\infty}^{\infty} \hat{a}_{n+N}^\dagger \hat{a}_n$ and $\chi = e^{j\theta}m/2$. Each term in the $\hat{T}_N$ operator represents the creation of a photon at mode $n+N$ resulting from the annihilation of another at mode $n$. It is immediate to check that $G_N(\chi,\varphi_b)$ is hermitian as required by the unitarity of $\hat{S}$. Furthermore, it is straightforward to prove that $[\hat{T}_N,\hat{T}_M]=[\hat{T}_N,\hat{N}_{ph}]=0$ so (18) can be expressed as:

$$\hat{S} = \exp(j\chi\hat{T}_N + j\chi^*\hat{T}_N^\dagger)\exp(j\varphi_b\hat{N}_{ph}) \equiv \exp(j\hat{Q})\exp(j\varphi_b\hat{N}_{ph}). \tag{19}$$

To prove that (19) is a faithful representation of the scattering operator it is enough to demonstrate that it verifies any of the transformations (16). For instance:

$$\hat{S}\hat{a}_n^\dagger\hat{S}^\dagger = e^{j\hat{Q}}e^{j\varphi_b\hat{N}_{ph}}\hat{a}_n^\dagger e^{-j\varphi_b\hat{N}_{ph}}e^{-j\hat{Q}} = \exp(j\,\mathsf{ad}_{\hat{Q}})\exp(j\varphi_b\,\mathsf{ad}_{\hat{N}_{ph}})(\hat{a}_n^\dagger), \tag{20}$$

where, for any pair of operators $\hat{A}$, $\hat{B}$, we define the adjoint action of $\hat{A}$ over $\hat{B}$, $\mathsf{ad}_{\hat{A}}(\hat{B})$, as $\mathsf{ad}_{\hat{A}}(\hat{B})=[\hat{A},\hat{B}]$. The inner part of (20) can be computed by use of the Baker-Campbell-Hausdorff lemma [13]:

$$\exp(j\varphi_b\,\mathsf{ad}_{\hat{N}_{ph}})(\hat{a}_n^\dagger) = \sum_{p=0}^\infty \frac{1}{p!}(j\varphi_b)^p\,\mathsf{ad}_{\hat{N}_{ph}}^{(p)}(\hat{a}_n^\dagger) = \sum_{p=0}^\infty \frac{1}{p!}(j\varphi_b)^p\,\hat{a}_n^\dagger = e^{j\varphi_b}\hat{a}_n^\dagger \tag{21}$$

where $\mathsf{ad}_{\hat{A}}^{(p)}(\hat{B}) = \overbrace{\mathsf{ad}_{\hat{A}}\,\ldots\,\mathsf{ad}_{\hat{A}}}^{p\text{ times}}(\hat{B})$ and we have used that $\mathsf{ad}_{\hat{N}_{ph}}^{(p)}(\hat{a}_n^\dagger) = \hat{a}_n^\dagger$. Now,

$$\exp(j\,\mathsf{ad}_{\hat{Q}})(\hat{a}_n^\dagger) = \sum_{p=0}^\infty \frac{j^p}{p!}\,\mathsf{ad}_{\hat{Q}}^{(p)}(\hat{a}_n^\dagger). \tag{22}$$

Using mathematical induction it is straightforward to prove that:

$$\mathsf{ad}_{\hat{Q}}^{(p)}(\hat{a}_n^\dagger) = \sum_{s=0}^p \binom{p}{s}\chi^s(\chi^*)^{p-s}\hat{a}_{n+(2s-p)N}^\dagger \tag{23}$$

and, hence, (20) is:

$$\hat{S}\hat{a}_n^\dagger \hat{S}^\dagger = e^{j\varphi_b} \sum_{p=0}^{\infty} \frac{j^p}{p!} \sum_{s=0}^{p} \binom{p}{s} \chi^s (\chi^*)^{p-s} \hat{a}_{n+(2s-p)N}^\dagger . \tag{24}$$

Figure 2 shows graphically the process of the coefficient construction for each resulting creation operator given by (24). We now must now show that (24) is identical to:

$$\hat{S}\hat{a}_n^\dagger \hat{S}^\dagger = \sum_{q=-\infty}^{\infty} C_q \hat{a}_{n+qN}^\dagger \tag{25}$$

with the coefficients given by (8). Lets assume that $q$ is odd (a similar line of reasoning can be followed for the case where $q$ is even) and consider all the terms that contribute to $\hat{a}_{n+qN}^\dagger$ for a fixed value of $q$. Comparing (25) and (26) then $q=2s-p$ and this means that $p=2s-q$ must be odd and, furthermore, $p \geq q$ since we have $s=(1/2)(p+q)$ and $0 \leq s \leq p$. Then we may write $p=q+2n$ where $n=0, 1, 2,...$ and hence $s=q+n$. The change of variables $n=p-s$, $q=2s-p$ applied to (24) gives:

$$\hat{S}\hat{a}_n^\dagger \hat{S}^\dagger = e^{j\varphi_b} \sum_{q=-\infty}^{\infty} \left[ j^q \sum_{n=0}^{\infty} \frac{(-1)^n}{(2n+q)!} \binom{q+2n}{q+n} \chi^{q+n}(\chi^*)^n \right] \hat{a}_{n+qN}^\dagger =$$
$$= e^{j\varphi_b} \sum_{q=-\infty}^{\infty} \left( \frac{jm}{2} e^{j\theta} \right)^q \left[ \sum_{n=0}^{\infty} \frac{(-1)^n}{(2n+q)!} \binom{q+2n}{q+n} \left(\frac{m}{2}\right)^{2n} \right] \hat{a}_{n+qN}^\dagger . \tag{26}$$

Now, recalling the series definition of the first kind Bessel functions:

$$J_q(m) = \left(\frac{m}{2}\right)^q \sum_{n=0}^{\infty} \frac{(-1)^n (m/2)^{2n}}{n!(q+n)!}, \tag{27}$$

we finally get:

$$\hat{S}\hat{a}_n^\dagger \hat{S}^\dagger = \sum_{q=-\infty}^{\infty} e^{j\varphi_b}(je^{j\theta})^q J_q(m) \hat{a}_{n+qN}^\dagger = \sum_{q=-\infty}^{\infty} C_q \hat{a}_{n+qN}^\dagger, \quad (28)$$

which completes the proof.

## 3.4 Diagrammatic computation of $\hat{S}\hat{a}_n^\dagger \hat{S}^\dagger$

In this subsection we show that (28) can be alternatively obtained by means of a diagrammatic approach based on the perturbative expansion of $\hat{S}$ in terms of the coupling $2|\chi| = m$. This approach will be used in the following section to compute the action of the new phase-modulation operator. First, and since the action of the $\exp(j\varphi_b \hat{N}_{ph})$ part of $\hat{S}$ in (19) is diagonal, we focus here on the $\exp(j\hat{Q})$ contribution, which is the operator responsible of mode coupling. In practice this amounts to set $\varphi_b = 0$ in (28). To illustrate the diagrammatic approach, figure 3 depicts a simple diagram representing the possible *transitions* starting from an input photon present at a mode characterized by a frequency number $n_0$ and a modulating frequency given by a number $N=1$. These transitions, represented by arrows in the diagram, are of two types: upconverting transitions where the photon is promoted from a mode with frequency number $n$ to a mode with frequency number $n+(N=1)=n+1$, and downconverting transitions where the photon is transferred from a mode with frequency number $n$ to a mode with frequency number $n-(N=1)=n-1$. Transitions begin and end in *nodes*. A *path* in the diagram is defined as any connected sequence of transitions (upconverting, downconverting or mixed) starting from the input node on the left vertex and ending in any other node. Allowed paths are only those which follow the arrows of the transitions.

The number of transitions in a path defines its perturbative *order*. Each path of order $k$ with, say, $n_u$ upconverting transitions and $n_d$ downconverting transitions ($n_u + n_d = k$) is also characterized by an *amplitude* defined as $(j\chi)^{n_u}(j\chi^*)^{n_d}/k!$. It is straightforward to show that there are $k!/n_u!n_d!$ different paths ending in the same node and thus having the same amplitude. It is also immediate to realize that if we attach each node with the number of paths ending in it, the resulting structure is that of a Tartaglia triangle since, for a given perturbative order $k$, the different numbers of the form $k!/n_u!n_d!$ with $n_u+n_d=k$ are precisely the coefficients of the expansion of the binomial $(a + b)^k$.

The diagrammatic method to compute the amplitude probability coefficients for a given input-output mode transition under the action of $\hat{S}$ is then described as follows: (a) first, for each possible group of paths with order $k$ connecting the given input and output modes, multiply the number of paths $k!/n_u!n_d!$ by their (common) amplitude $(j\chi)^{n_u}(j\chi^*)^{n_d}/k!$, and (b) add the contributions from all the different orders $k$.

For example, referring to figure 4, if we want to compute the amplitude probability coefficient for the $n_0 \rightarrow n_0$ transition, we first start with zero order paths (left part of the figure). There is only one with coefficient $1/0! = 1$. Then second order paths follow (there are no first-order paths) as seen in the upper right part of figure 5. Two different paths can be identified (shown in different colors in the figure). The total contribution for these two, second order paths is $-2|\chi|^2/2!$ The next step would be to compute the contribution of fourth order paths (there are no third order paths). The six possible options are shown in the lower right part of figure 4 and the total contribution is given by $6|\chi|^4/4!$ The procedure can be further continued to get:

$$A_{n_0 \to n_0 + 0N} = C_0 = \frac{1}{0!} - \frac{2|\chi|^2}{2!} + \frac{6|\chi|^4}{4!} + \cdots = \sum_{n=0}^{\infty} \frac{(-1)^n |\chi|^{2n}}{(n!)^2} = J_0(2|\chi|) = J_0(m). \tag{29}$$

Following, for instance, a similar procedure for the transition $n_o \to n_o + N$ we get:

$$A_{n_0 \to n_0 + 1N} = C_1 = je^\theta \left[ \frac{m}{2} - \frac{1}{2}\left(\frac{m}{2}\right)^3 + \cdots \right] = je^{j\theta} J_1(m). \tag{30}$$

In general, figure 3 shows that the transition amplitude for $n_0 \to n_0+qN$ with $q \geq 0$ only involves paths with orders of the form $k = q+2s$ with $s = 0, 1, 2, \ldots$ and the associated number of transitions are $n_u = q+s$ and $n_d = s$ (the case $q < 0$ can be analyzed similarly). Then, the total transition amplitude is:

$$A_{n_0 \to n_0 + qN} = C_q = \sum_{s=0}^{\infty} \frac{(j\chi)^{q+s}(j\chi^*)^s}{(q+2s)!} \frac{(q+2s)!}{(q+s)!s!} = (je^{j\theta})^q J_q(m) \tag{31}$$

which is consistent with (15) —with the $\varphi_b$ contribution omitted— and confirms the validity of the diagrammatic procedure.

## 4. Quantum Phase Modulation without Non-Positive Frequencies

### 4.1 Modified scattering operator

As mentioned before, eq. (13) involves, for sufficient high $q$, unphysical modes associated to zero or negative frequencies. However, the ratio of the optical carrier to the modulation frequency is of the order $\omega_0/\Omega = n_0/N \approx 10^4$ and, since the number of PM-generated sidebands is typically of the order of ten, low-frequency modes are never generated in practice. This situation is shown in the upper part of figure 5 where we plot a qualitative spectrum (the horizontal scale represents the mode number rather than the frequency) of a phase-modulated signal for an input single-mode state with mode number $n_0$ and $N = 1$. However, the formal quantum expression of $\hat{S}$ allows the existence and creation of photons whose mode index is not bounded as $n > 0$. For instance, the intermediate trace in figure 5 depicts the possibility of generating unphysical modes with $n \leq 0$, a situation that could be possible if the definition for $\hat{S}$ is kept as it stands in (18). The problem with this operator is simply that it allows an arbitrary decrease of the mode index. We solve this inconsistency by defining a different operator $\hat{S}$, whose form is similar to (18),

$$\hat{S} \equiv \hat{S}_N(\chi, \varphi_b) = \exp\left[j\hat{G}_N(\chi, \varphi_b)\right] = \exp\left[j(\chi\hat{T}_N + \chi^*\hat{T}_N^\dagger + \varphi_b\hat{N}_{ph})\right], \qquad (32)$$

but modify the operators in the exponent as:

$$\hat{T}_N = \sum_{m=1}^{\infty} \hat{a}_{m+N}^\dagger \hat{a}_m \qquad \hat{T}_N^\dagger = \sum_{m=1}^{\infty} \hat{a}_m^\dagger \hat{a}_{m+N} \qquad \hat{N}_{ph} = \hat{T}_0 = \sum_{m=1}^{\infty} \hat{a}_m^\dagger \hat{a}_m, \qquad (33)$$

which only involve positive-frequency modes while preserving the unitarity of $\hat{S}$. These modified operators provide equal probability of either rising or lowering the frequency of incoming photons, except for the case when lowering the incoming frequency would result in a zero or negative value. It is straightforward to check that $[\hat{S}, \hat{N}_{ph}] = 0$, and therefore $\hat{S}$ also conserves the number of photons.

Logically, the new definition must entail some reordering or modification of the transition amplitudes between modes, and these effects should be more pronounced when a certain transition involves modes which lay close to the bound $n=0$. This is shown qualitatively in the lower trace of figure 5, where we observe that modes with non-positive frequencies are no longer allowed. This modification will become explicit when we compute the expression of the unitary transformation of creation operators, $\hat{S}\hat{a}_n^\dagger \hat{S}^\dagger$, which represents the basic quantity in the quantum theory from which the transformation of arbitrary states can be derived.

## 4.2 Diagrammatic computation of $\hat{S}\hat{a}_n^\dagger \hat{S}^\dagger$

To obtain the transition amplitudes induced by the new operator $\hat{S}$ we modify the approach used in section 3.4. That diagrammatic procedure for deriving of the action of $\hat{S}$ simply sums, in terms of paths of different order, the probability amplitudes contributing to a given input-output transition. The possible paths are simply those generated by the exponential of the operator $\hat{G}_N$, which at every increasing order opens the possibility of new up- and downconverting transitions. When we consider the operator $\hat{S}$ instead of $\hat{S}$ two differences arise: first, only physical output modes with indices $n>0$ are allowed, but the resulting structure should still be that of a linear canonical transformation since $\hat{S}$ is the exponential of a quadratic operator composed of products $\hat{a}^\dagger \hat{a}$. This means that the sought-for transformation rule is of the form:

$$\hat{S}\hat{a}_{n_o}^\dagger \hat{S}^\dagger = \sum_{n>0} C_n(n_o) \hat{a}_n^\dagger \ , \qquad (34)$$

for a certain set of output modes $n$ and where the coefficients $C_n(n_o)$ may now depend on the initial mode index $n_0$. Secondly, when a path allowed for $\hat{S}$ arrives at a mode number $0 < n \leq N$ the possible paths can only continue upward because the operator $\hat{G}_N$ defined in (32) cannot create $n \leq 0$ modes since $[\hat{T}_N^\dagger, \hat{a}_n^\dagger] = 0$ for $0 < n \leq N$.

To illustrate this last difference we keep on with the previous example for which the modulating frequency is characterized by a number $N=1$. In this case, the process for downward photon creation is stopped when the photon arrives at the level characterized by $n=1$. This changes the path structure for $\hat{S}$ as compared to that of $\hat{\mathsf{S}}$. We illustrate this point in the upper right part of figure 6, where we plot some paths that contribute to the perturbative computation of the probability amplitude representing the transition to a final $n=1$ state. In solid trace we draw some of the allowed transitions while we display as well, in broken trace, some of the forbidden transitions when the modulator behavior is characterized by $\hat{S}$. Note that for $\hat{\mathsf{S}}$ all the transitions are allowed and, therefore all the displayed paths should be taken into account, whereas for $\hat{S}$ some of the paths do not contribute. The question is therefore how to subtract the contributions of the non-allowed paths from $\hat{\mathsf{S}}$ in order to get an expression for the action of $\hat{S}$. In this context we define a *forbidden path* as a path for $\hat{\mathsf{S}}$ which contains at least one forbidden transition, and therefore is not a path contributing to $\hat{S}$. For instance, in the lower right part of figure 6 we have outlined an example of such a forbidden path.

Although apparently formidable, the subtraction of the forbidden paths is simplified by the following observation. The lower left and right parts of figure 6 illustrate that for each forbidden path leading to a state characterized by $n>0$ there is another unique forbidden path that leads to a state $–n<0$ which is built from the original one by *transmitting* it through the line corresponding to $n=0$ and interchanging, from this point on, upward by downward transitions and vice versa. Since this property holds for each forbidden path and each path represents a perturbative contribution to the total transition amplitude, this property also holds for the perturbative sum over all forbidden paths. Then, the procedure to obtain the coefficients in (34) from the coefficients in expansion (28) is basically to subtract the $C_{-n-n_o}$ term which ends in $-n$ <0 to the coefficient $C_{n-n_o}$ which ends in $n$. There is, however, a difference. The contribution to the probability amplitude for each forbidden path differs from that of its equivalent transmitted (through $n=0$) path in that the transitions occurring below the $n=0$ line swap and hence, upconverting transitions are now affected by a $j\chi^*$ coefficient while downconverting transitions are characterized by a $j\chi$ coefficient. This means that, in the transmitted path, the phases $\theta$ corresponding to transitions that occur *after* reaching the mode $n=0$ for the first time, must change sign. For instance, paths ending in $n=-1$ require a single inversion of the phase:

$$C_{-1-n_o} = (je^{j\theta})^{-1-n_o} J_{-1-n_o}(m) = (je^{j\theta})^{-1}(je^{j\theta})^{-n_o} J_{-1-n_o}(m) \rightarrow$$
$$\rightarrow (je^{-j\theta})^{-1}(je^{j\theta})^{-n_o} J_{-1-n_o}(m) = (-1)(je^{j\theta})^{1-n_o} J_{-1-n_o}(m), \quad (35)$$

which is easily generalized to:

$$C_{-n-n_o} = (je^{j\theta})^{-n-n_o} J_{-n-n_o}(m) \rightarrow (-1)^n (je^{j\theta})^{n-n_o} J_{-n-n_o}(m). \quad (36)$$

This is the correct quantity to be subtracted to $C_{n-n_o} = (je^{j\theta})^{n-n_o} J_{n-n_o}(m)$. We finally arrive at the following equivalent solutions of (34), where the diagonal phase $\varphi_b$ has been reintroduced:

$$\hat{S}\hat{a}^\dagger_{n_o}\hat{S}^\dagger = \sum_{n=1}^{\infty} e^{j\varphi_b}(je^{j\theta})^{n-n_o}\left[J_{n-n_o}(m) - (-1)^n J_{-n-n_o}(m)\right]\hat{a}^\dagger_n =$$
$$= \sum_{n=1}^{\infty} e^{j\varphi_b}(je^{j\theta})^{n-n_o}\left[J_{n-n_o}(m) - (-1)^{n_o} J_{n+n_o}(m)\right]\hat{a}^\dagger_n = \quad (37)$$
$$= \sum_{s=1-n_o}^{\infty} e^{j\varphi_b}(je^{j\theta})^s\left[J_s(m) - (-1)^{n_o} J_{s+2n_o}(m)\right]\hat{a}^\dagger_{n_o+s} \equiv \sum_{q=1}^{\infty} D_{q,n_o} \hat{a}^\dagger_q$$

where we have defined the matrix elements $D_{q,n_o}$. This transformation, which has been derived for $N=1$, can be generalized to arbitrary modulation step $N$ as follows. Given $n_0$, the interaction with the phase modulator generates modes at $n_0 + sN$. If we decompose $n_0$ as $n_0 = q_0 N - r_0$ with $0 \leq r_0 < N$, the allowed values of $s$ run from $1 - q_0$ to $\infty$. Now, the coefficient describing the $s$-th generated sideband is the same as that in (37), since the diagrammatic structure resulting from $N > 1$ is similar to that of $N = 1$, the only differences being that (a) up- and downconverting transition now occur in steps of $N$, and (b) the first forbidden mode, where forbidden paths are to be swapped, has now an index $-r_0$. Then, using the expression in the third line of (37) and reordering the indices we get:

$$\hat{S}\hat{a}^\dagger_{n_o}\hat{S}^\dagger = \sum_{s=1-q_o}^{\infty} e^{j\varphi_b}(je^{j\theta})^s \left[ J_s(m) - (-1)^{q_o} J_{s+2q_o}(m) \right] \hat{a}^\dagger_{n_o+sN} \qquad (38)$$
$$= \sum_{q=1}^{\infty} e^{j\varphi_b}(je^{j\theta})^{q-q_o} \left[ J_{q-q_o}(m) - (-1)^{q_o} J_{q+q_o}(m) \right] \hat{a}^\dagger_{qN-r_0} = \sum_{q=1}^{\infty} D_{q,q_o} \hat{a}^\dagger_{qN-r_0},$$

which is the desired generalization. Eq. (37) is recovered by setting $N=1$, so that $q_0 = n_0$ and $r_0 = 0$. We stress that these expressions define a unitary theory because the operator $\hat{S}$ in (29) is unitary by construction. It is however illustrative to carry out the explicit check of unitarity starting from (37) or (38), which simply requires to check that, for positive $p_0$, $q_0$:

$$\sum_{q=1}^{\infty} D^\dagger_{p_0,q} D_{q,q_0} = \sum_{q=1}^{\infty} D^*_{q,p_0} D_{q,q_0} = \delta_{p_0,q_0}. \qquad (39)$$

Although straightforward, this calculation is rather lengthy and has been omitted.

We also point out that the modifications introduced by coefficient $D_{q,q_0}$ in (37) or (38) as compared with the classical coefficients $C_{q-n_o}$ are not significant at optical frequencies, so that the description of PM by means of operator $\hat{S}$ is consistent with the classical point of view. As mentioned before, in this case we have $\omega_0/\Omega = n_0/N \approx q_0 \approx 10^4$ and the number of significant sidebands generated by PM is typically of the order of ten. Then, in the first expression of (38), we have $s<10$ and $s+2q_0\sim 10^4$ which means that the second term inside the brackets is negligible.

Finally, to show that operator $\hat{S}$ behaves as qualitatively described in figure 5, and also to show that the classical picture holds for high-index modes and moderate number of sidebands, we have constructed an example. In figure 7 we have plotted the resulting quantum sidebands for a phase modulation with $N=1$ of a single photon with low mode number, $n_0=6$ (left), and with large mode number, $n_0=10$ (right). In both cases the modulation index is $m=5$. In (a) above, we show the quantities $|C_{q-n_o}|^2=|J_q(m)|^2$ which represent both the relative power between classical sidebands, and also the probability of a transition from mode $n_0$ to mode $n = n_0 + q$ if the scattering operator $\hat{S}$ is used. In (b) below, we show the analogous transition probabilities to the $q$-th sideband $|D_{q+n_o,n_o}|^2$ as given by (37). Focusing first in the left plot, in (a) the sidebands are symmetric around the input photon mode index, but the photon may jump to a negative frequency with finite probability. In (b) the scattering does not allow negative or zero frequencies, thus disturbing the symmetry of the sidebands. However, the disturbance is less pronounced for large values of the mode index $n$. In the second example on the right, the classical values in (a) provide a negligible probability of transition to a negative frequency. In (b) below we observe that the overall transition probabilities, including the symmetry of the sidebands, are not significantly changed by $\hat{S}$.

## 5. Quantum Multitone Phase Modulation

Up to this point we have considered the case where the input signal to the modulator electrode is a single sinusoidal tone as given by (4). In practice however, an important set of applications of the phase modulator is based on multitone signal modulation [4, 9], where a set of sinusoidal tones each one featuring a different frequency, phase and possibly amplitude is fed to the modulator electrode. It is thus of interest to extend the model developed so far to account for this important practical case. We first consider the simplest case, that is, when the input signal is composed of two different tones. We extend afterwards this model to arbitrary tone modulation under the assumption of low modulation index, which is always verified in practice.

When the input signal is composed of two modulating tones with amplitudes, phases and frequencies characterized by parameters $(m_1, \theta_1, N)$ and $(m_2, \theta_2, M)$, respectively, it is straightforward to extend the expression of the modulator scattering operator given by (29)-(30):

$$S_{(N,M)}\left(\varphi, \frac{m_1}{2}e^{j\theta_1}, \frac{m_2}{2}e^{j\theta_2}\right) = \exp(j\varphi\hat{N}_{ph}) \cdot \exp(\hat{X}+\hat{Y}) \;,$$

$$\hat{X} = j\frac{m_1}{2}e^{j\theta_1}T_N + j\frac{m_1}{2}e^{-j\theta_1}T_N^\dagger \;, \qquad (40)$$

$$\hat{Y} = j\frac{m_2}{2}e^{j\theta_2}T_M + j\frac{m_2}{2}e^{-j\theta_2}T_M^\dagger \;.$$

We now make use of the generalized Baker-Campbell-Hausdorff lema [13]:

$$\exp(\hat{X}+\hat{Y}) = \exp\left(-\frac{1}{2}[\hat{X},\hat{Y}] + \hat{O}(m^3)\right) \cdot \exp(\hat{X}) \cdot \exp(\hat{Y}) =$$

$$= \exp\left(-\frac{1}{2}[\hat{X},\hat{Y}] + \hat{O}(m^3)\right) \cdot \hat{S}_N\left(\frac{m_1}{2}e^{j\theta_1}\right) \cdot \hat{S}_M\left(\frac{m_2}{2}e^{j\theta_2}\right) \;, \qquad (41)$$

where $\hat{O}(m^3)$ is an operator affected by a factor proportional to the third power of the modulation index. Substituting (41) in (40) we get:

$$S_{(N,M)}\left(\varphi, \frac{m_1}{2}e^{j\theta_1}, \frac{m_2}{2}e^{j\theta_2}\right) =$$
$$= \exp(j\varphi\hat{N}_{ph}) \cdot \exp\left(-\frac{1}{2}[\hat{X},\hat{Y}] + \hat{O}(m^3)\right) \cdot \hat{S}_N\left(\frac{m_1}{2}e^{j\theta_1}\right) \cdot \hat{S}_M\left(\frac{m_2}{2}e^{j\theta_2}\right). \quad (42)$$

We now proceed to compute the transformation of an arbitrary creation operator:

$$S_{(N,M)} a_{n_o}^\dagger S_{(N,M)}^\dagger =$$
$$= \exp\left[j\varphi\hat{N}_{ph}\right]\exp\left[-\frac{1}{2}[\hat{X},\hat{Y}] + \hat{O}(m^3)\right]\hat{S}_N\hat{S}_M a_{n_o}^\dagger \hat{S}_M^\dagger \hat{S}_N^\dagger \exp\left[+\frac{1}{2}[\hat{X},\hat{Y}] - \hat{O}(m^3)\right]\exp\left[-j\varphi\hat{N}_{ph}\right] \quad (43)$$

Using (38) we have:

$$\hat{S}_M a_{n_o}^\dagger \hat{S}_M^\dagger = \sum_{q=1-q_o}^{\infty} \left(je^{j\theta}\right)^q \left[J_q(m_2) - (-1)^{q_o} J_{q+2q_o}(m_2)\right] \hat{a}_{n_o+qM}^\dagger = \sum_{q=1-q_o}^{\infty} D_q(q_o, m_2) \hat{a}_{n_o+qM}^\dagger$$

$$\hat{S}_N \hat{S}_M a_n^\dagger \hat{S}_M^\dagger \hat{S}_N^\dagger = \sum_{q=1-q_o}^{\infty} D_q(q_o, m_2) \hat{S}_N \hat{a}_{n_o+qM}^\dagger \hat{S}_N^\dagger = \{n_o + qM = s_o N + u\} = \quad (44)$$

$$= \sum_{q=1-q_o}^{\infty} D_q(q_o, m_2) \sum_{s=1-s_o}^{\infty} D_s(s_o, m_1) \hat{a}_{n_o+qM+sN}^\dagger = \sum_{q=1-q_o}^{\infty} \sum_{s=1-s_o}^{\infty} D_q(q_o, m_2) D_s(s_o, m_1) \hat{a}_{n_o+qM+sN}^\dagger .$$

Now, expanding:

$$\exp\left(\pm\frac{1}{2}[\hat{X},\hat{Y}] + \hat{O}(m^3)\right) \approx \hat{I} \pm \frac{1}{2}[\hat{X},\hat{Y}] + \hat{O}(m^3) \quad (45)$$

we finally arrive at:

$$S_{(N,M)} a_{n_o}^\dagger S_{(N,M)}^\dagger = e^{j\varphi_b} \hat{S}_N \hat{S}_M a_{n_o}^\dagger \hat{S}_M^\dagger \hat{S}_N^\dagger + \left[e^{j\varphi_b} \hat{S}_N \hat{S}_M a_{n_o}^\dagger \hat{S}_M^\dagger \hat{S}_N^\dagger, \hat{Q}(m^2)\right] \quad (46)$$

where $\hat{Q}(m^2)$ is an operator affected by a factor proportional to the second power of the modulation indices. Under practical operation conditions $m_i << 0$ (i=1,2) and therefore the second term in the right side of equation (46) can be neglected. In such a case:

$$\begin{aligned}
S_{(N,M)} \hat{a}_{n_o}^\dagger S_{(N,M)}^\dagger &\approx e^{j\varphi_b} \hat{S}_N \hat{S}_M \hat{a}_{n_o}^\dagger \hat{S}_M^\dagger \hat{S}_N^\dagger = e^{j\varphi_b} D_0(q_o, m_2) D_0(s_o, m_1) \hat{a}_{n_o}^\dagger + \\
&+ e^{j\varphi_b} D_0(q_o, m_2) D_1(s_o, m_1) \hat{a}_{n_o+N}^\dagger + e^{j\varphi_b} D_0(q_o, m_2) D_{-1}(s_o, m_1) \hat{a}_{n_o-N}^\dagger + \\
&+ e^{j\varphi_b} D_1(q_o, m_2) D_0(s_o, m_1) \hat{a}_{n_o+M}^\dagger + e^{j\varphi_b} D_{-1}(q_o, m_2) D_0(s_o, m_1) \hat{a}_{n_o-M}^\dagger + \ldots
\end{aligned} \quad (47)$$

Equation (47) illustrates the creation of the modulation sidebands around the optical carrier corresponding to each of the two modulating subcarriers, the remaining terms (not shown explicitly) corresponding to intermodulation products and harmonic distortion terms.

To generalize the previous result, we will now assume that the input signal to the modulator is a CW optical signal with frequency characterized by $n_o$ while the input signal to the modulator electrodes is a multitone radiofrecuency signal composed of $K$ sinusoids with amplitude, phase and frequency given by $(m_i, \theta_i, N_i)$, $i=1,2,\ldots K$. Since $n_o \gg N_i$, we can assume that $q_0, s_0, \cdots \to \infty$ and thus:

$$\begin{aligned}
&S_{(N_1, N_2, \cdots N_K)} \hat{a}_{n_o}^\dagger S_{(N_1, N_2, \cdots N_K)}^\dagger \approx \\
&\approx e^{j\varphi_b} \sum_{q_1=-\infty}^{\infty} \sum_{q_2=-\infty}^{\infty} \cdots \sum_{q_K=-\infty}^{\infty} \left(je^{j\theta_1}\right)^{q_1} \cdots \left(je^{j\theta_K}\right)^{q_K} J_{q_1}(m_1) \cdots J_{q_K}(m_K) \hat{a}_{n_o+q_1 N_1+\cdots q_K N_K}^\dagger = \\
&= e^{j\varphi_b} \left(\prod_{r=1}^{K} J_0(m_i)\right) \hat{a}_{n_o} + e^{j\varphi_b} \sum_{i=1}^{K} \left(\prod_{\substack{r=1 \\ r \neq i}}^{K} J_0(m_i)\right) \left[je^{j\theta_i} J_1(m_i) \hat{a}_{n_o+N_i}^\dagger - je^{-j\theta_i} J_{-1}(m_i) \hat{a}_{n_o-N_i}^\dagger\right] + \ldots
\end{aligned} \quad (48)$$

Again, the remaining terms (not shown explicitly) corresponding to intermodulation products and harmonic distortion terms. Of special interest in practice is the case where all modulation indexes are $\ll 1$. In this case, (48) can be further simplified by using $J_k(x) \approx x^k/k!$ to yield:

$$\begin{aligned}
&S_{(N_1, N_2, \ldots N_m)} \hat{a}_{n_o}^\dagger S_{(N_1, N_2, \ldots N_m)}^\dagger \approx \\
&= e^{j\varphi_b} \sum_{q_1=-\infty}^{\infty} \sum_{q_2=-\infty}^{\infty} \cdots \sum_{q_m=-\infty}^{\infty} \left(je^{j\theta_1}\right)^{q_1} \left(je^{j\theta_2}\right)^{q_2} \cdots \left(je^{j\theta_m}\right)^{q_m} \left(\frac{m_1^{q_1}}{q_1!}\right) \left(\frac{m_2^{q_2}}{q_2!}\right) \cdots \left(\frac{m_m^{q_m}}{q_m!}\right) \hat{a}_{n_o+q_1 N_1+q_2 N_2+\cdots q_m N_m}^\dagger = \\
&= e^{j\varphi_b} \hat{a}_{n_o} + e^{j\varphi_b} \sum_{k=1}^{m} m_k \left[je^{j\theta_k} \hat{a}_{n_o+N_k}^\dagger + je^{-j\theta_k} \hat{a}_{n_o-N_k}^\dagger\right] + \hat{Q}(m^2)
\end{aligned} \quad (49)$$

This formula represents the generalization of the single-tone, narrowband quantum phase modulation described in [1].

## 6. Summary and Conclusions

We have presented a detailed analysis of electro-optical phase modulation under quantum regime. Phase modulation has been described, for the first time to the best of our knowledge, in terms of a multimode unitary scattering operator defined on the Hilbert space of physical, positive-frequency modes, and shown to provide results consistent with the classical expressions when the relevant modes belong to optical bands. The model has been employed to characterize the important case of multitone radiofrequency modulation of an optical signal. This result is of interest in the characterization of the quantum properties of modulated radiation, and also in the design of quantum information systems employing modulation.

## Acknowledgements

This paper is supported by Ministerio de Ciencia y Tecnología, Spain, through Project TEC2008-02606 and through Quantum Optical Information Technology (QOIT), a CONSOLIDER-INGENIO 2010 Project. It is also funded by the Generalitat Valenciana through the PROMETEO 2008/092 research excellency award

# FIGURE CAPTIONS

**Figure 1:** (Upper) Typical configuration of a waveguide electro-optic phase modulator. (Lower) Black-Box representation of the phase modulator under quantum regime

**Figure 2:** Graphical representation of the process of coefficient construction for each resulting creation operator given by (24).

**Figure 3:** Simple diagram representing the possible *transitions* starting from an input photon present at a mode characterized by a frequency number $n_0$ and a modulating frequency given by a number $N=1$.

**Figure 4:** Auxiliary diagrams to compute the amplitude probability coefficient for the $n_0 \to n_0$ transition with $N=1$. Left: zero order path. Upper right: Second order paths. Lower right: Fourth order paths.

**Figure 5:** Qualitative spectrum (the horizontal scale represents the mode number rather than the frequency) of a phase-modulated signal for an input single-mode state with mode number $n_0$ and $N=1$.

**Figure 6:** (Upper left): Allowed and forbidden paths. (Upper right): a Second order forbidden path. (Lower): Diagram showing that for each forbidden path leading to a state characterized by $n>0$ (lower left) there is another unique forbidden path (lower right) that leads to a state $-n<0$ built from the original one by *transmitting* it through the line corresponding to $n=0$ and interchanging, from this point on, upward by downward transitions and vice versa

**Figure 7:** Transition probabilities of phase-modulated one-photon states at modes $n_0=6$ (left) and $n_0=10$ (right), see the text.

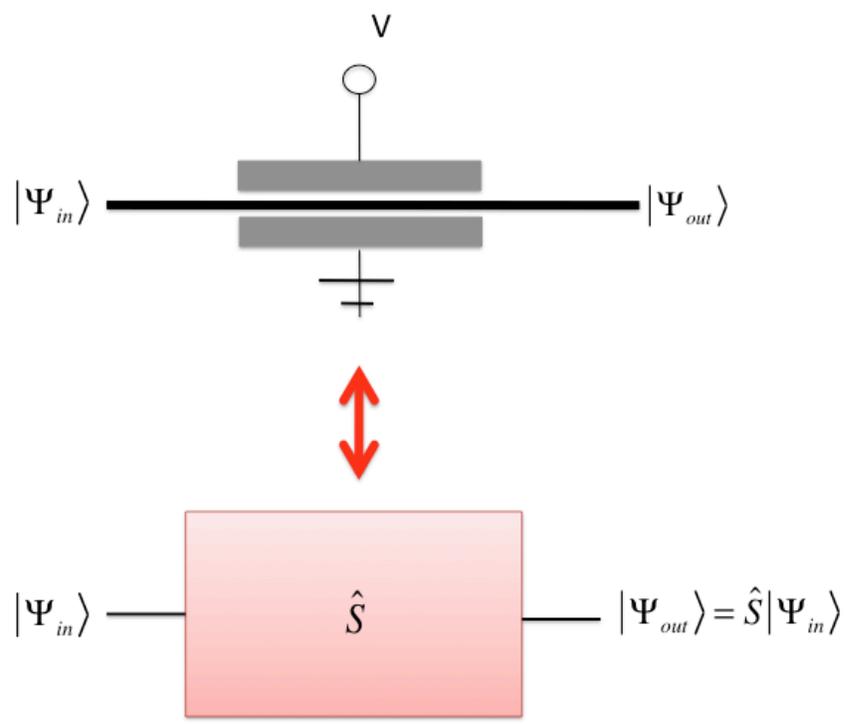

**FIGURE 1**

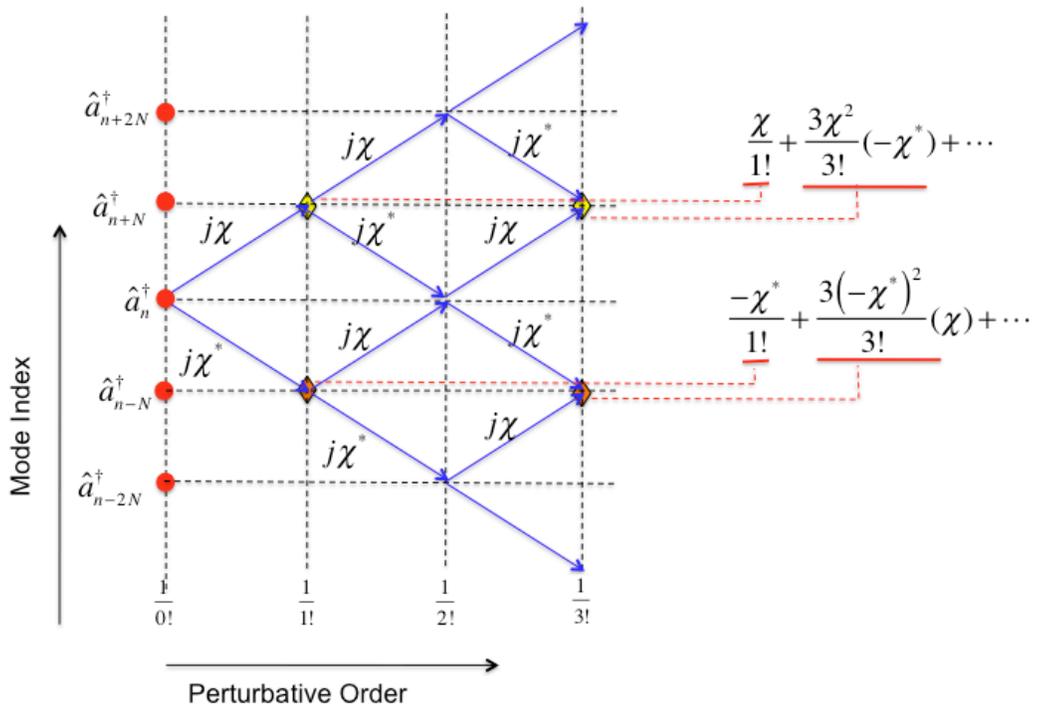

**FIGURE 2**

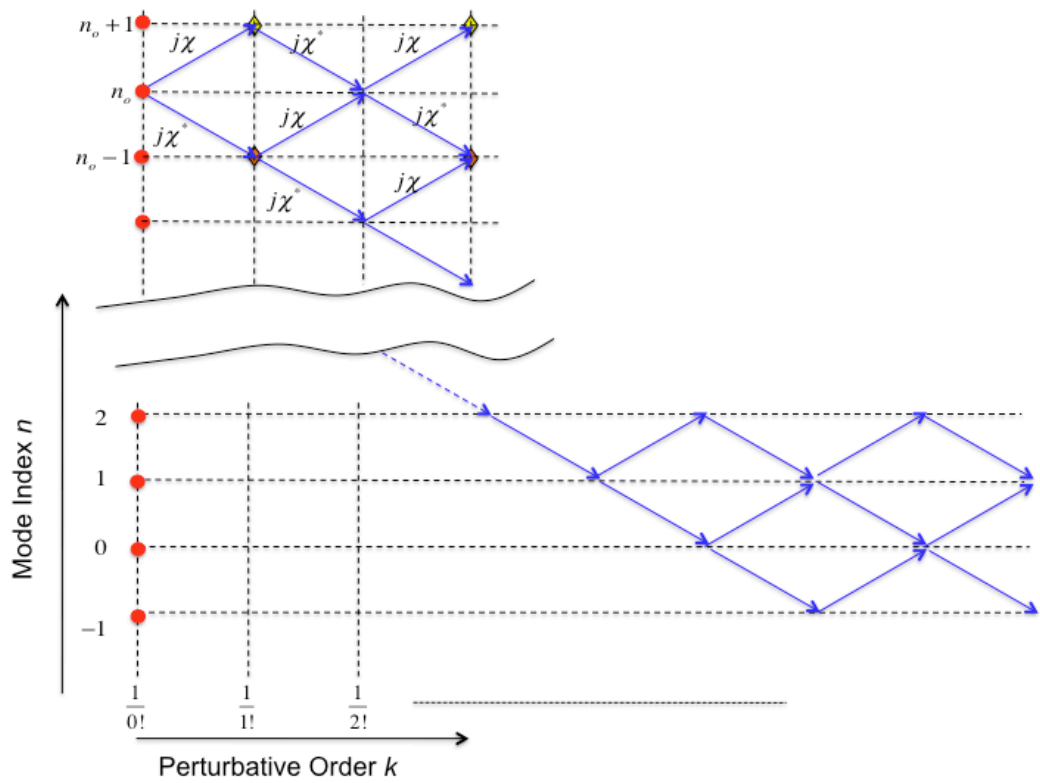

**FIGURE 3**

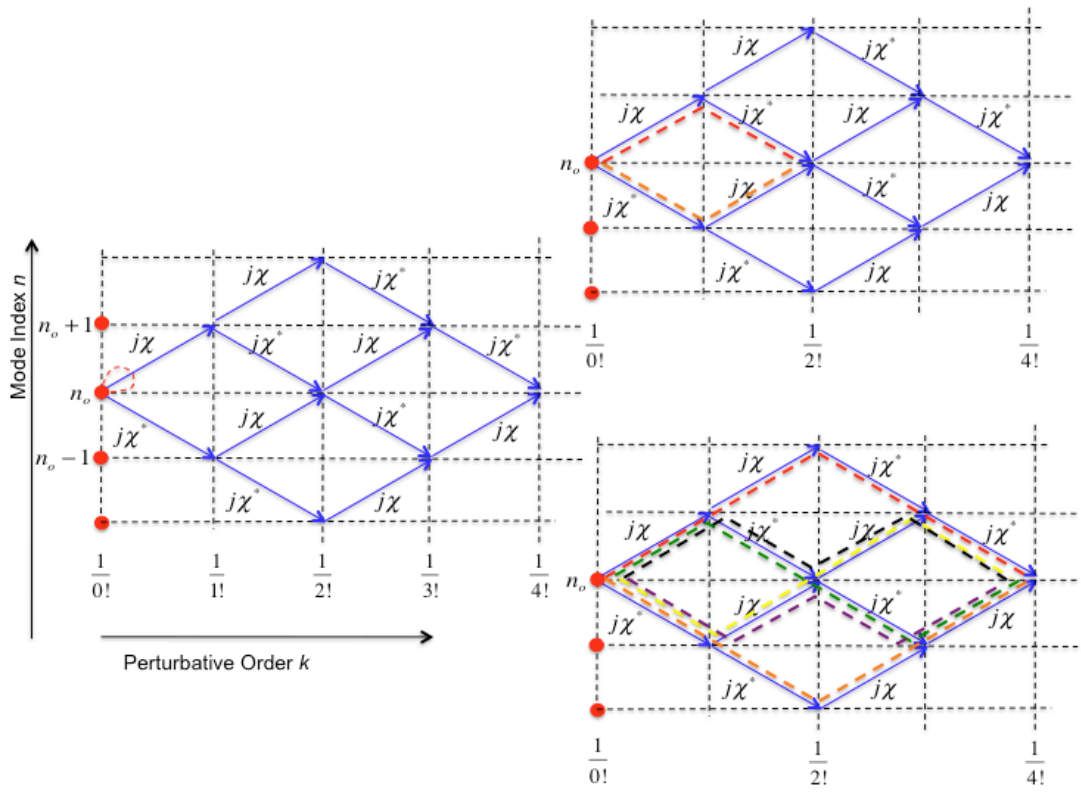

**FIGURE 4**

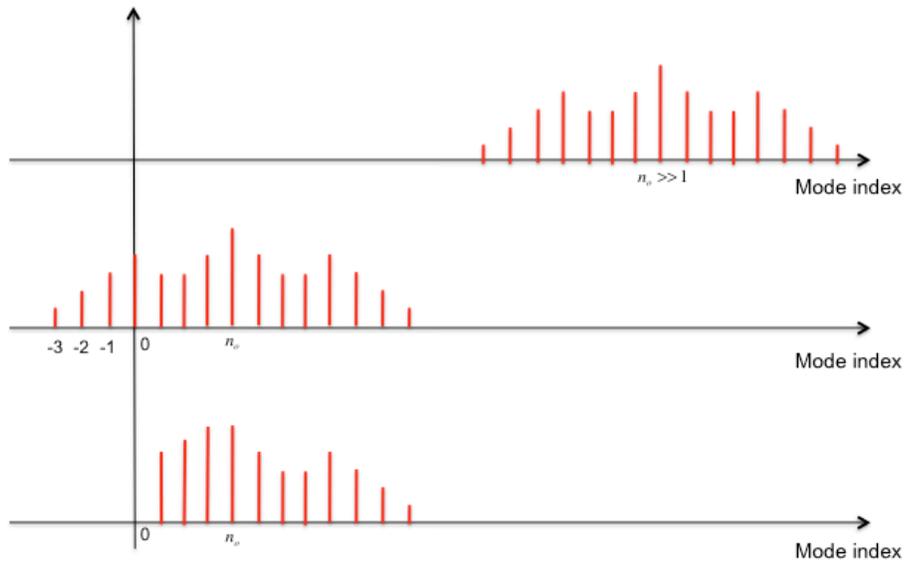

**FIGURE 5**

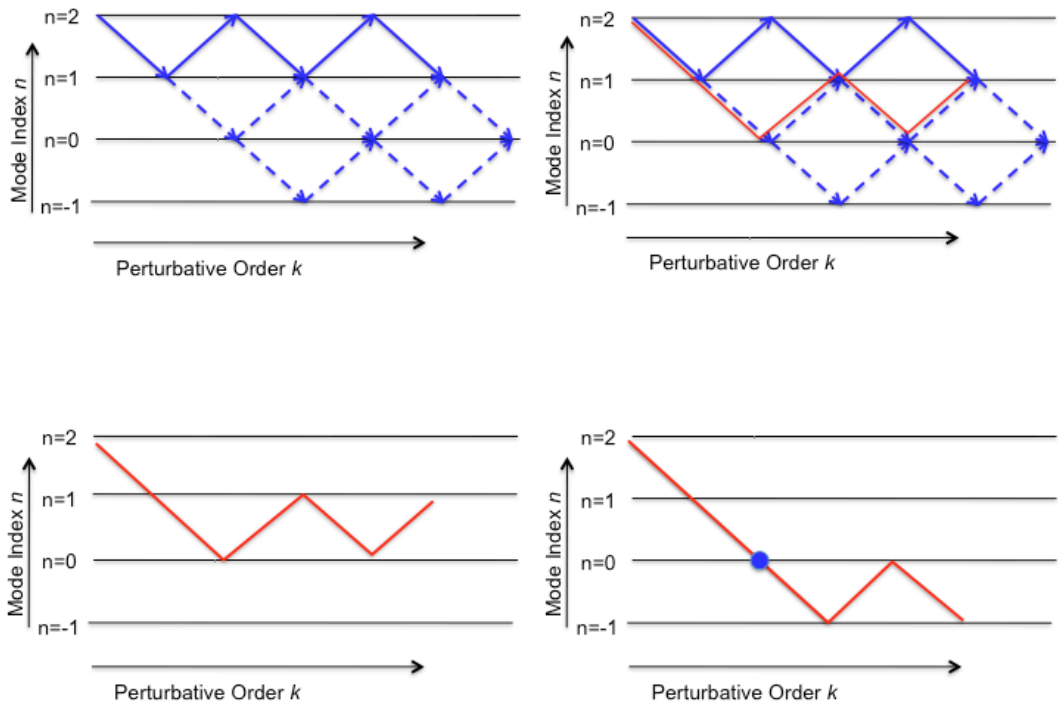

**FIGURE 6**

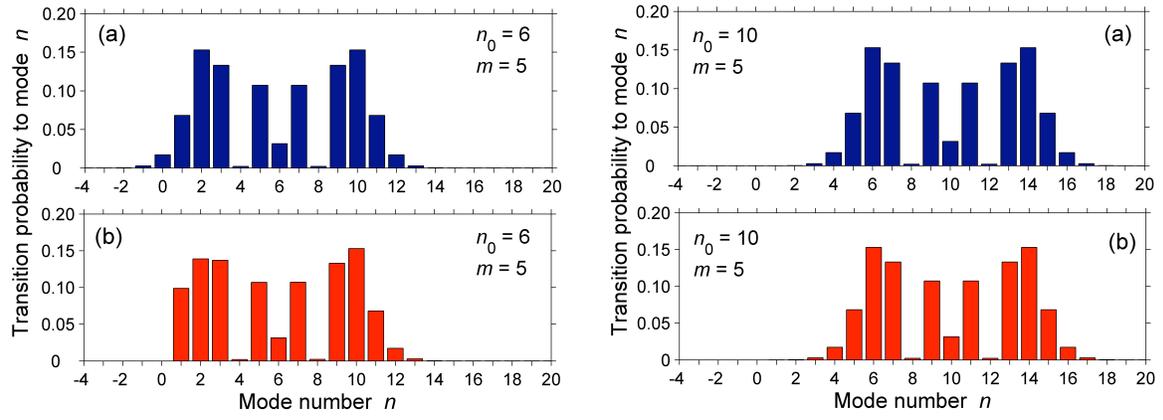

**FIGURE 7**